\documentclass[12pt]{article}

\usepackage{amssymb}
\usepackage{indentfirst}
\usepackage{amssymb,epsfig,amsmath,accents,setspace}
\usepackage{theorem}
\usepackage[usenames,dvipsnames]{xcolor}
\usepackage[active]{srcltx}
\usepackage{bm}
\usepackage{multirow}
\usepackage{lscape}
\usepackage{epstopdf}
\usepackage[justification=centering]{caption}

\usepackage{titlesec}

\usepackage{tikz}

\usepackage{url}
\usepackage[breaklinks]{hyperref}
\hypersetup{
    colorlinks=true,       
    linkcolor=red,          
    citecolor=blue,        
}

\usepackage[sort&compress]{natbib}
\bibpunct{(}{)}{;}{a}{,}{,}

\newtheorem{assumption}{Assumption}

\theoremstyle{plain} { \theorembodyfont{\rmfamily}

}

\newtheorem{corollary}{Corollary}
\newtheorem{theorem}{Theorem}

\newcommand{\bbR}{{\mathbb R}}

\newcommand{\prob}{\mathbb{P}}

\newcommand{\expect}{\mathbb{E}}

\def\halmos{\mbox{\quad$\square$}}
\def\Halmos{\mbox{\quad$\square$}}

\newenvironment{proof}{\vspace{1ex}\noindent{\em Proof.}\hspace{0.5em}}
    {\vspace{1ex}}

\newenvironment{pfof}[1]{\vspace{1ex}\noindent{\em Proof of #1.}\hspace{0.5em}}
    {\vspace{1ex}}

\usepackage{setspace}
 \numberwithin{equation}{section}

 \numberwithin{equation}{section}

\begin{document}
\title{Optimal Payoff under the Generalized Dual Theory of Choice
\thanks{The authors acknowledge financial support from the General Research Fund of the Research Grants Council of Hong Kong SAR (Project No. 14200917).}
}
\author{Xue Dong He\thanks{Corresponding Author. Room 505, William M.W. Mong Engineering Building, Department of Systems Engineering and Engineering Management, The Chinese University of Hong Kong, Shatin, N.T., Hong Kong, Telphone: +852-39438336, Email: xdhe@se.cuhk.edu.hk.} and Zhaoli Jiang\thanks{Department of Systems Engineering and Engineering Management, The Chinese University of Hong Kong, Shatin, N.T., Hong Kong, Email: zljiang@se.cuhk.edu.hk.}}
\maketitle

\begin{abstract}
We consider portfolio optimization under a preference model in a single-period, complete market. This preference model includes \citeauthor{YaariM:87dtc}'s dual theory of choice and quantile maximization as special cases. We characterize when the optimal solution exists and derive the optimal solution in closed form when it exists. The payoff of the optimal portfolio is a digital option: it yields an in-the-money payoff when the market is good and zero payoff otherwise. When the initial wealth increases, the set of good market scenarios remains unchanged while the payoff in these scenarios increases. Finally, we extend our portfolio optimization problem by imposing a dependence structure with a given benchmark payoff and derive similar results. 

\medskip



\end{abstract}

\section{Introduction}
Consider the following preference representation
\begin{align}\label{eq:GeneralizedYaari}
  V(X) = \int_{[0,1]} F_{ X}^{-1}(z)m(dz),
\end{align}
where $m$ is a probability measure on $[0,1]$ and $F_{X}^{-1}$ stands for the {\em right-continuous} quantile function of $X$, i.e., the right-continuous inverse of the cumulative distribution function (CDF) of $X$, with $F_{X}^{-1}(0):=\lim_{z\downarrow 0}F_{X}^{-1}(z)$ and $F_{X}^{-1}(1):=\lim_{z\uparrow 1}F_{X}^{-1}(z)$. By Lemma A.1 in \citet{GhossoubHe2020:ComparativeRA} and the integration-by-part formula, and defining $w(s):=\int_{(0,s]}m(dz),s\in [0,1]$, we can rewrite $V(X)$ for bounded $X$ as
\begin{align*}
  V(X) &= (1-w(1))F_X^{-1}(0)+ \int_{(0,1]}\Delta F_X^{-1}(z)dw(z)+ \int_{\mathbb{R}}x d[w(F_X(x))]\\
  & = (1-w(1))F_X^{-1}(0)+ \int_{(0,1]}\Delta F_X^{-1}(z)dw(z) -\int_{-\infty}^0 w(F_X(x))dx\\
  & + \int_0^{+\infty} \big(w(1)-w(F_X(x))\big)dx,
\end{align*}
where $F_X$ is the CDF of $X$ and $\Delta F_X^{-1}$ denotes the difference between $F_{X}^{-1}$ and its left-continuous version. Note from the above that when $w$ is continuous on $(0,1]$ and $w(1)=1$, i.e., when $m$ has no atom on $[0,1]$, $V(X)$ becomes the preference representation under the dual theory of choice \citet{YaariM:87dtc}. Therefore, we can consider \eqref{eq:GeneralizedYaari} to be a generalization of the dual theory of choice.

In the present paper, we study portfolio maximization under preferences \eqref{eq:GeneralizedYaari} in a single-period, complete market. This problem has been studied in \citet{HeXDZhouXY:2011PortfolioChoiceviaQuantiles} by assuming that $m$ has a density function and that a certain monotonicity condition holds. \citet{RuschendorfVanduffel2020:OnTheConstruction} also assume $m$ to have a density function, and they extend the result in \citet{HeXDZhouXY:2011PortfolioChoiceviaQuantiles} by removing the monotonicity condition imposed therein. The assumption of a density function for $m$, however, excludes many interesting preference models. For instance, \citet{Manski1988:Ordinal}, \citet{Chambers2009:Axiomatization}, \citet{rostek2010quantile}, \citet{deCastro2019:StaticDynamicQuantile} axiomatizes quantile maximization as a decision model; see also \citet{deCastro2019:PortfolioSelection}, \citet{Giovannetti2013:AssetPricing}, and \citet{deCastro2019:DynamicQuantileModels} for the applications of the quantile maximization model in portfolio selection and asset pricing. The $\alpha$-quantile of $X$ is a special case of $V(X)$ by setting $m$ to be a Dirac measure on $\alpha$, which does not admit a density function. On the other hand, in the context of risk management, \citet{KouPeng2016:OnTheMeasurement} propose the class of distortion risk measures, which take similar forms to the negative of \eqref{eq:GeneralizedYaari}. In this class, two particularly interesting risk measures are value-at-risk and median shortfall \citep{KouEtal2012:ExternalRiskMeasures}, and these two risk measures are special cases of \eqref{eq:GeneralizedYaari} by setting $m$ to be certain Dirac measures that do not have density functions.

In this paper, we derive the optimal solution to the portfolio maximization under preferences \eqref{eq:GeneralizedYaari} for a general probability measure $m$. More precisely, we completely characterizes when the problem has an optimal solution. When the optimal solution exists, we show that is unique and the payoff of the optimal portfolio is a digital option: it leads to certain winning payoff in some market scenarios and to zero otherwise. We then extend our model to the setting in \citet{BernardEtal2014:ExplicitRepresentationCostEfficient,BernardEtal2014:OptimalClaims}, and \citet{BernardEtal2015:OptimalPayoffs}, where the agent focuses on payoffs that have a particular dependence structure, modeled by a copula, with a given benchmark payoff. We derive similar results in this extended model.

Our solution method relies on the so-called quantile approach to convert the portfolio optimization problem into a problem of finding the optimal quantile function. For this approach, see for instance \citet{Schied2004:OnTheNeymanPearsonProblem}, \citet{CarlierGDanaR:06licu,CarlierDana:2008TwoPersonsEfficientRiskSharing,CarlierDana2011:OptimalDemandForContigentClaims}, \citet{JinHQZhouXY:08bps}, \citet{BernardTian:2009OptimalReinsuranceTailRiskMeasures}, \citet{HeXDZhouXY:2011PortfolioChoiceviaQuantiles}, \citet{BernardEtal2014:OptimalClaims}, and \citet{BernardEtal2015:OptimalPayoffs}. The new technical contribution of the paper is to solve the optimal quantile problem that results from portfolio maximization under preferences \eqref{eq:GeneralizedYaari} because the existing methods, such as the ones in \citet{XiaZhou2012:ArrowDebreuEquilibriaRDU}, \citet{Xu2014:ANote}, and \citet{RuschendorfVanduffel2020:OnTheConstruction} cannot apply. Our idea of solving the optimal quantile problem originates from \citet{HeJinZhou2014:PortfolioSelectionRiskMeasure}, but our technical analysis differs from the latter due to different model setup. Moreover, we also consider the case in which the agent focuses on payoffs that have a particular dependence structure with a given benchmark payoff while the latter did not.


\section{Optimal Payoff under the Generalized Dual Theory of Choice}\label{se:StaticProblem}
Consider the following portfolio selection problem faced by an agent:
\begin{align}\label{eq:QuantileOptimization}
  \begin{array}{cl}
    \underset{X}{\text{Max}} & V(X) = \int_{[0,1]} F_{ X}^{-1}(z)m(dz)\\
    \text{Subject to} & \expect[\rho  X]\le  x,\quad  X\ge 0,
  \end{array}
\end{align}
where $\rho$, a positive, integrable random variable, represents the pricing kernel in a single-period complete market, $m$ is a probability measure on $[0,1]$, and $x>0$ represents the agent's initial wealth. The pricing kernel $\rho$ and payoffs $X$ are assumed to live on a probability space $(\Omega,{\cal F},\prob)$. Denote $\delta:=\expect[\rho]$.

%
%

Suppose the following assumption holds throughout this Section:
\begin{assumption}\label{as:NoAtom}
  $\rho$ has no atom, i.e., its CDF is continuous.
\end{assumption}

Denote $\underline{\rho}:=\text{essinf}\,\rho\ge 0$. Because $\rho>0$ almost surely, we must have $F_{\rho}^{-1}(z)>0,z\in (0,1)$. Following \citet{HeXDZhouXY:2011PortfolioChoiceviaQuantiles}, we reformulate problem \eqref{eq:QuantileOptimization} as
\begin{align}\label{eq:QuantileOptimizationQuantile}
  \begin{array}{cl}
    \underset{G(\cdot)\in \mathbb{G}}{\text{Max}} & \int_{[0,1]} G(z)m(dz)\\
    \text{Subject to} & \int_{0}^1F_{\rho}^{-1}(1-z)G(z)dz\le  x,\quad G(0)\ge 0,
  \end{array}
\end{align}
where $\mathbb{G}$ is the set of right-continuous quantile functions, i.e.,
\begin{align*}
  \mathbb{G}:=\{G(\cdot):[0,1]\rightarrow \mathbb{R}\cup\{\pm\infty\}|G(\cdot)\text{ is finite-valued, right-continuous,}\\
  \text{and increasing in }(0,1)\text{ and }G(0) = \lim_{z\downarrow 0}G(z),G(1) = \lim_{z\uparrow 1}G(z)\}.
\end{align*}
By the quantile approach \citep{Schied2004:OnTheNeymanPearsonProblem,CarlierGDanaR:06licu,CarlierDana:2008TwoPersonsEfficientRiskSharing,CarlierDana2011:OptimalDemandForContigentClaims,JinHQZhouXY:08bps,BernardTian:2009OptimalReinsuranceTailRiskMeasures,HeXDZhouXY:2011PortfolioChoiceviaQuantiles}, problems \eqref{eq:QuantileOptimization} and \eqref{eq:QuantileOptimizationQuantile} share the same optimal value and are equivalent in terms of the existence and uniqueness of the optimal solution. Furthermore, if $G^*(\cdot)$ is optimal to \eqref{eq:QuantileOptimizationQuantile}, then $G^*(1-F_\rho(\rho))$ is optimal to \eqref{eq:QuantileOptimization}. Therefore, we only need to solve problem \eqref{eq:QuantileOptimizationQuantile}.

Note that problems \eqref{eq:QuantileOptimization} and \eqref{eq:QuantileOptimizationQuantile} still share the same optimal value and are equivalent in terms of the existence of the optimal solution if Assumption \ref{as:NoAtom} is weakened to the one that the probability space $(\Omega,{\cal F},\prob)$ is atomless; see for instance \citet{Xu2014:ANewCharacterization}. The optimal solution of problem \eqref{eq:QuantileOptimization}, however, is not unique even if the optimal solution of \eqref{eq:QuantileOptimizationQuantile} is. For this reason and also for the sake of simplicity, we chose to impose Assumption \ref{as:NoAtom}.

Note that the objective function of \eqref{eq:QuantileOptimizationQuantile} is linear in the decision variable $G(\cdot)$. Therefore, we only need to optimize over the extremal points of the set of feasible quantiles to problem \eqref{eq:QuantileOptimizationQuantile}. \citet[Proposition D.3]{JinHQZhouXY:08bps} show that the extremal points are contained in the set of binary quantile functions
\begin{align*}
  \mathbb{S}:=\Big\{G(\cdot)|G(z)  = a+k\mathbf 1_{z\in[c,1]} \text{ for some }c\in(0,1)\text{ and }a,k\ge 0\\
  \text{such that } \int_{0}^1F_{\rho}^{-1}(1-z)G(z)dz\le  x\Big\}.
\end{align*}
Thus, we only need to consider
\begin{align}\label{eq:QuantileOptimizationQuantileSimple}
  \begin{array}{cl}
    \underset{G(\cdot)\in \mathbb{S}}{\text{Max}} & \int_{[0,1]} G(z)m(dz).
  \end{array}
\end{align}
Indeed, we can show that problems \eqref{eq:QuantileOptimizationQuantile} and \eqref{eq:QuantileOptimizationQuantileSimple} have the same optimal value and, consequently, if $G^*(\cdot)$ is optimal to problem \eqref{eq:QuantileOptimizationQuantileSimple}, then it is also optimal to \eqref{eq:QuantileOptimizationQuantile}.

Denote $f(a,k,c)$ as the objective function of problem \eqref{eq:QuantileOptimizationQuantileSimple} with $G(z)  = a+k\mathbf 1_{z\in[c,1]}$. Straightforward calculation shows that $f(a,k,c) = a +km([c,1])$. As a result, problem \eqref{eq:QuantileOptimizationQuantileSimple} can be written as
\begin{align}\label{eq:QuantileOptimizationQuantileTwo}
  \begin{array}{cl}
    \underset{a,k,c}{\text{Max}} & a +k m([c,1])\\
    \text{Subject to} & a\delta +k \int_{c}^1F_{\rho}^{-1}(1-z)dz\le x,\; a,k\ge 0,\; c\in (0,1).
  \end{array}
\end{align}
It is obvious that the optimal $k$ must bind the first constraint in the above. Consequently, we only need to maximize
\begin{align*}
  f(a,c):= \zeta(c)  x + [1-\delta\zeta(c)]a
\end{align*}
over $c\in(0,1)$ and $a\in[0, x/\delta]$, where
\begin{align*}
\zeta(c):=\frac{m([c,1])}{\int_{c}^1F_{\rho}^{-1}(1-z)dz}.
\end{align*}
One can see that
\begin{align*}
  \max_{a\in[0, x/\delta]} f(a,c) = \zeta(c)  x + \max(1-\delta\zeta(c),0) \frac{ x}{\delta} = \max(1/\delta,\zeta(c)) x.
\end{align*}
Therefore, we only need to maximize $\max(1/\delta,\zeta(c))$ over $c\in(0,1)$.

In the following, denote
\begin{align*}
  \gamma^*:=\sup_{c\in(0,1)}\zeta(c).
\end{align*}
Then, we have $\sup_{c\in(0,1)}\max(1/\delta,\zeta(c)) = \max(1/\delta,\gamma^*)$.

\begin{theorem}\label{th:MainTheorem}
The optimal value of problem \eqref{eq:QuantileOptimization} is $\max(1/\delta,\gamma^*) x$. Furthermore,
  \begin{enumerate}
   \item[(i)] If $\gamma^*=+\infty$, there exist $\beta_n\rightarrow \underline{\rho}$ such that $X_n:=k_n\mathbf 1_{\rho \le \beta_n}$ with $k_n:=x/\expect[\rho \mathbf 1_{\rho \le \beta_n}]$ is feasible and approaches the infinite optimal value when $n$ goes to infinity.
    \item[(ii)] If $\gamma^*\le 1/\delta$, $ X^*\equiv  x/\delta$ is optimal to problem \eqref{eq:QuantileOptimization}.
    \item[(iii)] If $1/\delta<\gamma^*=\sup_{c\in(0,1)}\zeta(c)<+\infty$ and the supremum is attained at $c^*\in(0,1)$, then $X^*:= k^*\mathbf 1_{\rho \le \beta^*}$
    with $\beta^*:=F^{-1}_\rho(1-c^*)$ and $k^*:= \frac{ x }{\expect[\rho \mathbf 1_{\rho \le \beta^*}]}$
    is optimal to problem \eqref{eq:QuantileOptimization}.
    \item[(iv)] If $1/\delta<\gamma^*=\sup_{c\in(0,1)}\zeta(c)<+\infty$ and the supremum is not attainable, then the optimal solution to problem \eqref{eq:QuantileOptimization} does not exist. Furthermore, for any $\beta_n:=F^{-1}_\rho(1-c_n)$ with $\zeta(c_n)\rightarrow \gamma^*$ as $n\rightarrow \infty$, $X_n:=k_n\mathbf 1_{\rho \le \beta_n}$ with $k_n:=x/\expect[\rho \mathbf 1_{\rho \le \beta_n}]$ is feasible and approaches the optimal value when $n$ goes to infinity.
  \end{enumerate}
\end{theorem}
\begin{proof}
Problem \eqref{eq:QuantileOptimizationQuantileTwo} has optimal value $\max(1/\delta,\gamma^*) x$, so does problem \eqref{eq:QuantileOptimization}. Consequently, in case (i), the optimal value of problem \eqref{eq:QuantileOptimization} is infinite. Furthermore, there exist $c_n$ such that $\zeta(c_n)\rightarrow \gamma^*=+\infty$. Because $m([c,1])\le 1$ for any $c\in[0,1]$ and $\int_{c}^1F_\rho^{-1}(1-z)dz$ is continuous, positive, and strictly decreasing in $c\in[0,1)$, $\zeta(c)$ is bounded in $[0,1-\delta]$ for any $\delta>0$. Consequently, we must have $c_n\rightarrow 1$. Define $\beta_n:=F^{-1}_\rho(1-c_n)$ and $X_n:=k_n\mathbf 1_{\rho \le \beta_n}$. It is straightforward to see that $X_n$ is feasible and approaches the infinite optimal value.

  In case (ii), the optimal value of problem \eqref{eq:QuantileOptimization} is $x/\delta$. It is obvious that $X^* = x/\delta$ is feasible and achieves this optimal value.

  In case (iii), the optimizer $c^*$ of $\zeta(\cdot)$, $a^*:=0$, and $k^*:=\frac{ x}{\int_{c^*}^1F_\rho^{-1}(1-z)dz}$ are optimal to problem \eqref{eq:QuantileOptimizationQuantileTwo}. As a result, $G^*(z) = a^*+k^*\mathbf 1_{z\in[c^*,1]}$ is optimal to problem \eqref{eq:QuantileOptimizationQuantile}, so $X^* = G^*(1-F_\rho(\rho))$ is optimal to problem \eqref{eq:QuantileOptimization}.

 In case (iv), the optimal value of problem \eqref{eq:QuantileOptimizationQuantile} is $x\gamma^*$.
We show that the optimal solution to problem \eqref{eq:QuantileOptimizationQuantile} does not exist. Suppose $G(\cdot)$ is optimal to problem \eqref{eq:QuantileOptimizationQuantile}. Then, $G(\cdot)$ cannot be a constant on $(0,1)$. Otherwise, suppose $G(z) = a,z\in(0,1)$, which also implies $G(0)= G(0+)=a$ and $G(1) = G(1-) = a$.
 Then,
 \begin{align*}
   \int_{[0,1]}G(z)m(dz)   = a \le  x/\delta< x \gamma^*,
 \end{align*}
 where the first inequality is the case because $G(\cdot)$ must satisfy the budget constraint. Thus, we conclude that $G(\cdot)$ cannot be a constant on $(0,1)$.

 Denote $\nu(\cdot)$ as the measure induced by $G(\cdot)$ on $(0,1]$, i.e., $\nu((0,z]):=G(z)-G(0)$ for any $z\in(0,1]$.  Because $G(\cdot)$ is nonconstant on $(0,1)$, $\nu(\cdot)$ has nonzero measure on $(0,1)$.
 Then, we have
 \begin{align*}
   \int_{[0,1]}G(z)m(dz) &=  G(0) + \int_{[0,1]}\int_{(0,z]}\nu(ds)m(dz)  = G(0) + \int_{(0,1]}\int_{[s,1]}m(dz)\nu(ds)\\
   & = G(0) + \int_{(0,1]}m([s,1])\nu(ds) = G(0) + \int_{(0,1]}\zeta(s)\int_s^{1}F_\rho^{-1}(1-z)dz \nu(ds)\\
   & < G(0) + \gamma^* \int_{(0,1]}\int_s^{1}F_\rho^{-1}(1-z)dz\nu(ds)\\
   & = G(0) +\gamma^*\left(\int_{0}^1F_\rho^{-1}(1-z)G(z)dz - G(0) \int_{0}^1F_\rho^{-1}(1-z)dz \right)\\
   &\le \gamma^* x,
 \end{align*}
 where the first inequality is the case because $\zeta(s)<\gamma^*,s\in(0,1)$ and $\nu$ has nonzero measure on $(0,1)$, and the second inequality is the case because $\int_{0}^1F_\rho^{-1}(1-z)G(z)dz\le x$ and $\gamma^*>1/\delta=\frac{1}{  \int_{0}^1F_\rho^{-1}(1-z)dz } $. Therefore, $G(\cdot)$ cannot attain the optimal value, so the optimal solution to problem \eqref{eq:QuantileOptimizationQuantile} does not exist. Consequently, problem \eqref{eq:QuantileOptimization} does not admit optimal solutions either.

 Finally, it is straightforward to see that $X_n:=k_n\mathbf 1_{\rho \le \beta_n}$ is feasible and approaches the optimal value as $n\rightarrow+\infty$.\Halmos
\end{proof}

We have completely solved problem \eqref{eq:QuantileOptimization} and the solution depends critically on the quantity $\gamma^*$. When $\gamma^*=+\infty$, the optimal value is infinite, so problem \eqref{eq:QuantileOptimization} is ill-posed. This result generalizes Theorem 3.4 in \citet{HeXDZhouXY:2011PortfolioChoiceviaQuantiles} and Theorem 4.1 in \citet{RuschendorfVanduffel2020:OnTheConstruction}. Furthermore, the asymptotically optimal strategy taken by the agent in this case is $X_n = k_n\mathbf 1_{\{\rho\le \beta_n\}}$ with $\beta_n\rightarrow \underline{\rho}$. In this strategy, the probability of having nonzero payoffs is nearly zero, so the strategy is extremely risky. When $\gamma^*\le 1/\delta$, the optimal payoff is constant, indicating the strategy of investing all money in the risk-free account, an extremely conservative strategy. When $\gamma^*=\sup_{c\in(0,1)}\zeta(c)>1/\delta$ and the supremum is not attainable, problem \eqref{eq:QuantileOptimization} is also ill-posed and the asymptotically optimal strategy is $X_n = k_n\mathbf 1_{\{\rho\le \beta_n\}}$ with $\beta_n:=F^{-1}_\rho(1-c_n)$ and $\zeta(c_n)\rightarrow \gamma^*$. If $m([c,1])$ is continuous in $[0,1]$, which is the case when $m$ admits a density, then $\zeta(c)$ is continuous in $[0,1)$. Consequently, we must have $c_n\rightarrow 1$ and thus $\beta_n\rightarrow \underline{\rho}$. As a result, the probability of $X_n$ taking nonzero payoffs is nearly zero, showing that it is an extremely risky strategy. Finally, when $\gamma^* = \zeta(c^*)\in(1/\delta,+\infty)$ for some $c^*\in(0,1)$, the optimal payoff is a digital option: receiving $k^*$ in good market scenarios ($\rho \le \beta^*$) or receiving nothing in bad market scenarios ($\rho >\beta^*$). 

Typically, $\gamma^*>1/\delta$, so the optimal solution, if exists, must be a digital option. Indeed, if $m$ has zero measure in a neighbourhood of $0$, e.g., in $[0,\epsilon]$ for some $\epsilon>0$, then
\begin{align*}
  \gamma^*\ge \zeta(\epsilon) = m([\epsilon,1])/\int_{\epsilon}^1F_\rho^{-1}(1-z)dz > m([\epsilon,1])/\int_{0}^1F_\rho^{-1}(1-z)dz > 1/\delta.
\end{align*}
 Therefore, $\gamma^*\le 1/\delta$ only when the agent imposes significant weight on the very left tail of the payoff, i.e., only when the agent is so conservative that she wants to minimize the maximum potential loss in the downside.

 When $1/\delta<\gamma^*=\zeta(c^*)<+\infty$ for some $c^*\in(0,1)$, the optimal solution is a digital option and is different from the optimal payoff of an expected utility maximizer which is smooth in the pricing kernel $\rho$. Note that when the initial wealth $x$ increases, $\beta^*$ does not change, so the probability that the digital option is ``in the money" at the terminal time does not change. On the other hand, the payoff $k^*$ of the digital option when it is in the money increases. This feature is different from the optimal payoff of a goal-reaching agent \citep{BrowneS:99rg}. Indeed, the optimal payoff of a goad-reaching agent is also a digital option. However, the probability that this option is in the money increases with respect to the initial wealth but the in-the-money payoff does not depend on the initial wealth. Similar observations are also made in \citet{HeXDZhouXY:2011PortfolioChoiceviaQuantiles}.
 
 Finally, we present the optimal payoff under quantile maximization:

\begin{corollary}\label{coro:QuantilePreCommitted}
  Suppose that $V(X)$ is the $\alpha$-quantile of $X$, i.e., that $m$ is the point mass at $\alpha\in(0,1)$. Then, $\gamma^* = \zeta(\alpha)>1/\delta$ and $ X^*:= k^*\mathbf 1_{\rho \le \beta^*}$
    with $\beta^*:=F^{-1}_\rho(1-\alpha)$ and $k^*:= \frac{ x }{\expect[\rho \mathbf 1_{\rho \le \beta^*}]}$
    is optimal to problem \eqref{eq:QuantileOptimization}.
\end{corollary}
\begin{pfof}{Corollary \ref{coro:QuantilePreCommitted}}
  Straightforward calculation shows that
  \begin{align*}
  \zeta(c) = \frac{1}{\int_{c}^1F_\rho^{-1}(1-z)dz }\mathbf 1_{\{c\le  \alpha\}},\quad c\in (0,1).
\end{align*}
Clearly,
\begin{align*}
   \gamma^*=\zeta(\alpha)=\frac{1}{\int_{\alpha}^1F_\rho^{-1}(1-z)dz }>\frac{1}{\int_{0}^1F_\rho^{-1}(1-z)dz }=1/\delta.
\end{align*}
As a result, Theorem \ref{th:MainTheorem}-(iii) applies and the proof completes.\halmos
\end{pfof}

\section{Optimal Payoffs with State-Dependent Constraints}
Consider the setting in \citet{BernardEtal2014:ExplicitRepresentationCostEfficient,BernardEtal2014:OptimalClaims}, and \citet{BernardEtal2015:OptimalPayoffs}: There is a benchmark payoff $A$ and the agent wants to choose payoff $X$ that has a given dependence structure with $A$, and this dependence structure is specified by a given copula function $C$; i.e., $X$ and $A$ follow the joint distribution:
\begin{align}\label{eq:CopulaConstraint}
  \prob(X\le x, A\le a) = C(F_X(x),F_A(a)),\quad x,a\in \bbR.
\end{align}
As a result, the problem faced by the agent is
\begin{align}\label{eq:QuantileOptimizationWithCopula}
  \begin{array}{cl}
    \underset{X}{\text{Max}} & V(X) = \int_{[0,1]} F_{X}^{-1}(z)m(dz)\\
    \text{Subject to} & \expect[\rho X]\le x,\quad X\ge 0,\\
    & \eqref{eq:CopulaConstraint}\text{ holds}.
  \end{array}
\end{align}
We impose the following assumption throughout this section:
\begin{assumption}\label{as:ConditioinalPricingKernel}
  (i) The conditional distribution of $\rho$ given $A$ is continuous. (ii) The distribution of $A$ is continuous.
\end{assumption}

We follow the approach in \citet{BernardEtal2014:OptimalClaims} to convert \eqref{eq:QuantileOptimizationWithCopula} into a problem of finding the optimal quantile. More precisely, similar to the quantile approach, we first find the payoff $X$ that minimizes the cost $\expect[\rho X]$ given the marginal distribution $F_X$ of $X$ and the copula constraint \eqref{eq:CopulaConstraint}. Theorem 3.1 in \citet{BernardEtal2014:OptimalClaims} shows that $X$ attains the minimum cost if and only if it is anti-comonotone with respect to $\rho$ conditional on $A$. With Assumption \ref{as:ConditioinalPricingKernel}-(i), $X$ is uniquely determined as $X = F_{X|A}^{-1}(1-F_{\rho|A}(\rho,A),A)$, where for any two random variables $Y_1$ and $Y_2$, we denote by $F_{Y_1|Y_2}(\cdot,y_2)$ and $F_{Y_1|Y_2}^{-1}(\cdot,y_2)$, respectively, the conditional CDF and right-continuous quantile function of $Y_1$ given that $Y_2=y_2$. Note that $F_{\rho|A}(\rho,A)$ is uniformly distributed and independent of $A$. 
The minimal cost is $\expect\left[\rho F_{X|A}^{-1}(1-F_{\rho|A}(\rho,A),A)\right]$.
Following \citet{BernardEtal2014:OptimalClaims}, we define
\begin{align}\label{eq:ZTransform}
  Z:= C^{-1}_{1|2}(1-F_{\rho|A}(\rho,A),F_A(A)),
\end{align}
where $C_{1|2}(\cdot,v)$ and $C^{-1}_{1|2}(\cdot, v)$ denote, respectively, the conditional CDF and right-continuous quantile function of the first component in the copula $C$, given that the second component takes value $v$. Then, $Z$ is uniformly distributed and $F^{-1}_{X|A}(1-F_{\rho|A}(\rho,A),A) = F_X^{-1}(Z)$ by Assumption \ref{as:ConditioinalPricingKernel}-(ii). Consequently, the minimal cost can be written as
\begin{align*}
  \expect\left[\rho F_X^{-1}(Z)\right] = \expect[\varphi(Z)F_X^{-1}(Z)] = \int_0^1\varphi(z)F_X^{-1}(z)dz,
\end{align*}
where
\begin{align}\label{eq:CondExRho}
  \varphi(z):=\expect[\rho\mid Z=z],\quad z\in (0,1).
\end{align}
Then, problem \eqref{eq:QuantileOptimizationWithCopula} can be transformed into
\begin{align}\label{eq:QuantileOptimizationQuantileWithCopula}
  \begin{array}{cl}
    \underset{G(\cdot)\in \mathbb{G}}{\text{Max}} & \int_{[0,1]} G(z)m(dz)\\
    \text{Subject to} & \int_{0}^1\varphi(z)G(z)dz\le x,\quad G(0)\ge 0.
  \end{array}
\end{align}
More precisely, problems \eqref{eq:QuantileOptimizationWithCopula} and \eqref{eq:QuantileOptimizationQuantileWithCopula} share the same optimal value and are equivalent in terms of the existence and uniqueness of the optimal solution. Furthermore, if $G^*(\cdot)$ is optimal to \eqref{eq:QuantileOptimizationQuantileWithCopula}, then $G^*(Z)$ is optimal to \eqref{eq:QuantileOptimizationWithCopula}. Therefore, we only need to solve problem \eqref{eq:QuantileOptimizationQuantileWithCopula}.

Note that \eqref{eq:QuantileOptimizationQuantile} becomes \eqref{eq:QuantileOptimizationQuantileWithCopula} if $F^{-1}_\rho(1-z)$ is replaced by $\varphi(z)$. In addition, $\varphi(z)>0,z\in (0,1)$ because $\rho$ is positive, and we have $\int_0^1\varphi(z)dz = \expect[\varphi(Z)]=\expect[\rho]=\delta$. Therefore, the solution to \eqref{eq:QuantileOptimizationQuantile} as derived in Section \ref{se:StaticProblem} leads to the solution to \eqref{eq:QuantileOptimizationQuantileWithCopula} and, consequently, to the solution to \eqref{eq:QuantileOptimizationWithCopula} immediately; see the following:

\begin{theorem}\label{th:MainTheorem}
The optimal value of problem \eqref{eq:QuantileOptimizationWithCopula} is $\max(1/\delta,\gamma^*) x$, where
\begin{align*}
\tilde \zeta(c):=\frac{m([c,1])}{\int_{c}^1\varphi(z)dz},\; c\in (0,1),\quad \tilde \gamma^*:=\sup_{c\in(0,1)}\tilde \zeta(c).
\end{align*}
Furthermore,
  \begin{enumerate}
   \item[(i)] If $\tilde \gamma^*=+\infty$, there exist $c_n\rightarrow 1$ such that $X_n:=k_n\mathbf 1_{Z \ge  c_n}$ with $k_n:= x/\expect[\rho \mathbf 1_{Z \ge  c_n}]$ is feasible and approaches the infinite optimal value as $n$ goes to infinity.
    \item[(ii)] If $\tilde \gamma^*\le 1/\delta$, $ X^*\equiv  x/\delta$ is optimal to problem \eqref{eq:QuantileOptimizationWithCopula}.
    \item[(iii)] If $1/ \delta<\tilde \gamma^*=\sup_{c\in(0,1)}\tilde \zeta(c)<+\infty$ and the supremum is attained at $c^*\in(0,1)$, then $X^*:= k^*\mathbf 1_{Z \ge c^*}$ with $k^*:= \frac{x }{\expect[\rho \mathbf 1_{Z \ge c^*}]}$
    is the unique optimal solution to problem \eqref{eq:QuantileOptimizationWithCopula}.
    \item[(iv)] If $1/ \delta<\tilde \gamma^*=\sup_{c\in(0,1)}\tilde \zeta(c)<+\infty$ and the supremum is not attainable, then the optimal solution to problem \eqref{eq:QuantileOptimizationWithCopula} does not exist. Furthermore, for any $c_n\in (0,1)$ with $\tilde \zeta(c_n)\rightarrow \tilde \gamma^*$ as $n\rightarrow \infty$, $ X_n:=k_n\mathbf 1_{Z \ge  c_n}$ with $k_n:= x/\expect[\rho \mathbf 1_{Z \ge c_n}]$ is feasible and approaches the optimal value as $n$ goes to infinity.
  \end{enumerate}
\end{theorem}

\begin{corollary}\label{coro:QuantilePreCommitted}
  Suppose that $V(X)$ is the $\alpha$-quantile of $X$, i.e., that $m$ is the point mass at $\alpha\in(0,1)$. Then, $\tilde \gamma^* = \tilde \zeta(\alpha)>1/\delta$ and $X^*:= k^*\mathbf 1_{Z\ge \alpha}$
    with $k^*:= \frac{ x }{\expect[\rho \mathbf 1_{Z\ge \alpha}]}$
    is optimal to problem \eqref{eq:QuantileOptimizationWithCopula}.
\end{corollary}

%
%



\section{Conclusions}
In this paper we considered portfolio optimization in a single-period, complete market. The agent's preferences are represented by a model that includes both \citeauthor{YaariM:87dtc}'s dual theory of choice and quantile maximization as special cases. Applying the quantile approach in the literature, we converted the portfolio optimization problem into a problem of finding the optimal quantile function. By solving the latter problem, we obtained a characterization of when the optimal portfolio exists and derived the optimal portfolio in closed form when it exists. It turns out that the payoff of the optimal portfolio is a digital option. Moreover, when the agent's initial wealth increases, the scenarios in which the digital option is in the money remain unchanged while the payoff in these scenarios increases. Finally, we extended our model to a setting in which the agent focuses on payoffs that are correlated with a given benchmark by a certain copula, and we derived similar portfolio optimization results.


\end{document}